\documentclass{iopart}

\usepackage[dvips]{graphicx}

\newcommand{\NF}{\textit{Nucl. Fusion} }
\newcommand{\PP}{\textit{Phys. Plasmas} }

\newcommand{\PFB}{\textit{Phys. Fluids B} }

\begin{document}

\title{Equilibrium reconstruction for Single Helical Axis reversed field pinch plasmas}
\author{E~Martines, R~Lorenzini, B~Momo, D~Terranova, P~Zanca, A~Alfier, F~Bonomo, A~Canton, A~Fassina, P~Franz, P~Innocente}
\address{Consorzio RFX, Associazione Euratom-ENEA sulla Fusione, 
         corso Stati Uniti 4, 35127 Padova, Italy}
\ead{emilio.martines@igi.cnr.it}

\begin{abstract}
Single Helical Axis (SHAx) configurations are emerging as the natural state for high current reversed field pinch (RFP) plasmas. These states feature the presence of transport barriers in the core plasma. Here we present a method for computing the equilibrium magnetic surfaces for these states in the force-free approximation, which has been implemented in the SHEq code. The method is based on the superposition of a zeroth order axisymmetric equilibrium and of a first order helical perturbation computed according to Newcomb's equation supplemented with edge magnetic field measurements. The mapping of the measured electron temperature profiles, soft X-ray emission and interferometric density measurements on the computed magnetic surfaces demonstrates the quality of the equilibrium reconstruction. The procedure for computing flux surface averages is illustrated, and applied to the evaluation of the thermal conductivity profile. The consistency of the evaluated equilibria with Ohm's law is also discussed.
\end{abstract}


\submitto{\PPCF}

\section{Introduction}
\label{intro}
The reversed field pinch (RFP) configuration has been known for a long time as a possible candidate for the magnetic confinement of fusion plasmas \cite{rfp1,rfp2}. Among toroidal configurations, it is the only one which could in principle achieve reactor-relevant conditions without the need of additional heating systems, thanks to its high level of ohmic power dissipation. Nevertheless, up to recent years the RFP was regarded as having poor confinement properties, due to the simultaneous presence of several MHD tearing modes with m=1 poloidal periodicity \cite{locking}. These modes are resonant at different radii within the plasma core, and their magnetic islands overlap, so that magnetic surfaces are destroyed and a high level of magnetic chaos ensues. As a consequence, plasmas with flat density and temperature profiles over the whole plasma core are produced in this condition, which is dubbed Multiple Helicity (MH) state. The presence of these modes is intrinsic to the configuration, since they are required to drive poloidal currents in the outer plasma region while the applied electric field is only toroidal (dynamo effect) \cite{dinamo1,dinamo2}. In the past, chaos reduction was obtained transiently by inductive poloidal current drive \cite{ppcd_mst,ppcd_rfx}, but a way of obtaining this effect in a stationary fashion is still lacking.

This view of the RFP as a stochastic plasma with bad confinement properties is radically changing, thanks to the recent discovery of a new class of equilibria, which have been dubbed Single Helical Axis (SHAx) states \cite{lorenzini_prl, np}. SHAx states are the result of a natural tendency of RFP plasmas to move, as dissipation changes, towards a single helicity (SH) condition, that is a condition where only one of the core-resonant tearing modes provides the dynamo effect \cite{sh0,sh1,sh2,sh3}. This tendency, which has been predicted since several years by 3D MHD simulations \cite{cappello_prl}, has been confirmed by experimental results in several RFP devices \cite{qsh_tpe,qsh_t2,qsh_mst,qsh_rfx}, where states in which the innermost resonant m=1 mode dominates over the others (secondary modes) have been observed. Such conditions are generally called Quasi Single Helicity (QSH) states, since they differ from the theoretical SH ones due to the fact that the secondary modes still have a non-negligible amplitude.
In particular, in the RFX-mod device (R = 2 m, a = 0.459 m) it has been observed, when feedback control of the radial magnetic field at the edge is applied \cite{martini,marrelli}, that the duration and frequency of occurrence of QSH states increases with plasma current \cite{np,carraro,piovesan}. The secondary mode amplitude, normalized to the total magnetic field, is also found to decrease with increasing plasma current, so that experimental QSH states are progressively approaching the ideal SH condition \cite{np,carraro,piovesan}. Plasma current increase is strongly correlated to an increase of electron temperature \cite{valisa}.

In the QSH states the dominant mode amplitude is found to increase with plasma current, inducing a bifurcation to a new topological structure of the magnetic field \cite{lorenzini_prl}. This happens because, beyond a threshold in the dominant mode amplitude, the X-point of the magnetic island separatrix collapses on the main magnetic axis, and the two disappear, leaving the original island O-point as the only magnetic axis. Thus, a helical plasma column is obtained in the axisymmetric device. These SHAx states, which represent a special flavour of the more general QSH condition, are more resilient to magnetic chaos, due to the disappearance of the separatrix \cite{escande_prl}. They constitute a new paradigm for high performance RFP plasmas, which could lead to a re-evaluation of the potential of this configuration for fusion reactor development.

In this paper we propose a method to determine the flux surface shape of RFP discharges in a SHAx state, which involves only the solution of ordinary differential equations. The method, which has been implemented in a code named SHEq (Single Helicity Equilibrium) starts from the symplectic representation of the magnetic field \cite{boozer_review},
\begin{equation}
\mathbf{B} = \nabla F\times \nabla\theta - \nabla\Psi\times\nabla\phi
\label{B_canonical}
\end{equation}
where $(\rho,\theta,\phi)$ is a generic system of curvilinear coordinates used to describe the toroidal system. $F\equiv F(\rho,\theta,\phi)$ and $\Psi\equiv \Psi(\rho,\theta,\phi)$ represent the toroidal flux through a constant $F$ surface and the poloidal flux outside a constant $\Psi$ surface, respectively, both divided by $2 \pi$. The existence of magnetic flux surfaces is assured if a function $\rho(\mathbf{x})$ such that
\begin{equation}
\mathbf{B} \cdot \nabla \rho = 0
\end{equation}
exists.
Flux surfaces are then described by the equation $\rho=const.$ (or $f(\rho)=const.$ for any $f$ function of $\rho$ only). The magnetic axis is defined by $\nabla \rho=0$. In the case of an axisymmetric plasma with circular cross-section the function $\rho$ can be taken to coincide with the radius $r$ of the magnetic flux surfaces, and $F(r)$ and $\Psi(r)$ may be interpreted as the usual toroidal and poloidal flux, respectively. In the case of a general magnetic field, $F(r,\theta,\phi)$ and $\Psi(r,\theta,\phi)$ in (1) are not constant over flux surfaces, flux surface existence being not guaranteed in the general 3D case. In other terms, we can say that adding a generic perturbation field to the axisymmetric one, the circular flux surfaces are destroyed, and it is not clear a priori if some other flux surfaces exist, and which can be the function $\rho$ that labels them. 

If one assumes a helical symmetry of the perturbation, so that $\Psi$ and $F$ depend only on $r$ and on $u=m\theta-n\phi$, the symmetry ensures flux surface existence, and it is straightforward to show that the helical flux $\chi(r,u)=m\Psi(r,u)-n F(r,u)$ is a flux function, that is $\mathbf{B}\cdot\nabla\chi=0$. Thus, the knowledge of $\chi(r,u)$ enables to display the shape of the magnetic surfaces. This is the basic hypothesis adopted to treat SHAx states.

The method that we propose for computing the helical flux in SHAx states is based on the approach described in ref. \cite{zt} for calculating the tearing mode eigenfunctions in a force-free RFP plasma in toroidal geometry. The SHAx states are therefore considered as composed of a dominant, saturated tearing mode superposed on an axisymmetric equilibrium. The tearing mode amplitude is assumed to be small with respect to the axisymmetric fields, so that a perturbative approach can be adopted. The same coordinate systems and the same notation of ref. \cite{zt} will be used.
The most relevant results of ref. \cite{zt} are reviewed in the two sections next, so as to give a self-consistent treatment of the equilibrium problem, the reader is referred to that paper for further details.

This paper is organized as follows: 
in Section \ref{zeroth} the zeroth-order axisymmetric equilibrium calculation is introduced; 
in Section \ref{first} the first order correction due to the dominant mode is computed; 
in Section \ref{helical} the method used to compute the flux surfaces is described and validated by studying the mapping of electron temperature, density soft X-ray emission measurements on them; 
in Section \ref{averaging} a method for computing flux surface averages is described, and an example application to power balance computation in a SHAx condition is given;
Section \ref{ohmic} is devoted to a discussion of the constraint given by the flux surface averaged Ohm's law parallel component.
Finally, in Section \ref{conclusions} conclusions are drawn.

\section{Zeroth-order equilibrium}
\label{zeroth}
Let us consider a zeroth-order axisymmetric toroidal plasma with circular cross-section, formed in a vacuum chamber with major radius $R_0$ and minor radius $a$. The flux surfaces are non-concentric circles, each having radius $r$, being horizontally shifted by a quantity $\Delta(r)$. The shift of the outermost flux surface is imposed as boundary condition (in the experiments this is obtained from external magnetic measurements). A point lying on one of these flux surfaces is identified by the radius $r$ of the surface, by the poloidal angle $\theta$ measured with respect to the inboard mid plane, and by the toroidal angle $\phi$. These coordinates, which we call geometric coordinates, $u^i$, are related to the standard cylindrical system $(R,\phi, Z)$ used to describe toroidal fusion devices by
\begin{eqnarray}
R = R_0-r\cos\theta+\Delta(r) \label{geo1} \\
Z = r\sin\theta.  \label{geo2}
\end{eqnarray}
where $R_0$ is the torus major radius.
The $(r, \theta, \phi)$ coordinate system is curvilinear and non-orthogonal, in order to properly take into account the toroidal geometry. A complete knowledge of the metric tensor is essential (see Appendix A for a brief reminder on curvilinear coordinates). The metric tensor and the Jacobian of the geometric coordinates are given in Appendix B.

The contravariant representation of the zeroth-order magnetic field associated with the geometric coordinate system is
\begin{equation}
{\bf B}_0 = {\bf\nabla}F_0(r)\times{\bf\nabla}\theta - {\bf\nabla}\Psi_0(r)\times{\bf\nabla}\phi +
{\bf\nabla}r\times{\bf\nabla}\nu(r,\theta)   
\end{equation}
where $F_0$ and $\Psi_0$ are, respectively, the toroidal and poloidal flux divided by $2\pi$. The equilibrium is fully defined once $F_0(r)$, $\Psi_0(r)$, $\Delta(r)$ and $\nu(r,\theta)$ are known.

Following the standard procedure for introducing flux coordinates \cite{dhaeseleer}, one can define a new poloidal angle as
\begin{equation}
\vartheta = \theta + \lambda(r,\theta)
\label{vartheta}
\end{equation}
with $\lambda(r,\theta)=\nu(r,\theta)/F'_0(r)$. Here and in the following a prime designates derivative with respect to $r$ of quantities which are functions of $r$ only. In the $w^i=(r,\vartheta,\phi)$ system, which we call flux coordinates, 
the magnetic field lines are straight
and the magnetic field contravariant representation is simply
\begin{equation}
{\bf B}_0 = {\bf\nabla}F_0(r)\times{\bf\nabla}\vartheta - {\bf\nabla}\Psi_0(r)\times{\bf\nabla}\phi.
\label{B0}
\end{equation}
This provides simple formulas for the contravariant components $B^i$, given by
\begin{equation}
B_0^r = 0 \qquad
B_0^\vartheta = \frac{1}{\sqrt{g_w}}\Psi'_0 \qquad
B_0^\phi = \frac{1}{\sqrt{g_w}}F'_0
\label{B_up}
\end{equation}
where $\sqrt{g_w}$ is the Jacobian of the flux coordinate system:
\begin{eqnarray}
\sqrt{g_w} = \left( \nabla r \cdot \nabla \vartheta \times \nabla \phi \right)^{-1}.
\label{sqrt_g}
\end{eqnarray}
The metric tensor and the Jacobian of the flux coordinates are also given in Appendix B.

The determination of the parameter $\lambda(r, \theta)$ in (\ref{vartheta}) is possible for a large aspect ratio torus following a perturbative approach. Amp\'ere's law allows to deduce the current density contravariant components $J^i$.
From the force balance condition and $B_0^r=0$ one gets $J_0^r=0$. Using this information, and performing an expansion in the small aspect ratio parameter $\epsilon=a/R_0$, it is possible to compute the quantity relating $\theta$ and $\vartheta$ as
\begin{equation}
\lambda(r,\theta) = \lambda_1(r)\sin\theta + \lambda_2(r)\sin 2\theta + o(\epsilon^3).
\label{lambda}
\end{equation}
where
\begin{equation}
\lambda_1(r) = \frac{r}{R_0}-\Delta'(r) \qquad
\lambda_2(r) = \frac{r}{4R_0}\lambda_1(r).
\end{equation}
The inverse of transformation (\ref{vartheta}) is then easily derived as
\begin{equation}
\theta = \vartheta - \lambda_1\sin\vartheta -
\left(\lambda_2-\frac{\lambda_1^2}{2}\right)\sin 2\vartheta + o(\epsilon^3).
\end{equation}
Using equation (\ref{lambda}) the relation between cylindrical $(R, \phi, Z)$ and flux coordinates $w^i$ can also be found:
\begin{eqnarray}
\fl
R = R_0 - r\cos\vartheta + \Delta(r) - r\lambda_1(r)\sin^2\vartheta +
\left(\frac{3}{2}r\lambda_1^2-2r\lambda_2\right)\sin^2\vartheta\cos\vartheta + o(\epsilon^3b)
\label{R}
\end{eqnarray}
\begin{eqnarray}
\fl
Z = r\sin\vartheta - \frac{r}{2}\lambda_1(r)\sin 2\vartheta +
\left(\frac{3}{2}r\lambda_1^2-2r\lambda_2\right)\sin\vartheta\cos^2\vartheta -
\frac{r}{2}\lambda_1^2(r)\sin\vartheta + o(\epsilon^3b)
\label{Z}
\end{eqnarray}
so $R \equiv R(r, \vartheta)$ and $Z \equiv Z(r, \vartheta)$.

The flux coordinate Jacobian for a large aspect ratio torus is, as stated  in Appendix B,
\begin{equation}
\sqrt{g_w} = \frac{R^2}{K(r)}
\label{jacob0}
\end{equation}
with
\begin{equation}
K(r) = \frac{R_0}{r}
\left(1+\frac{\Delta}{R_0}+\frac{r}{2R_0}\Delta'-\frac{r^2}{2R_0^2}+o(\epsilon^3)\right).
\label{K_r}
\end{equation}
It is possible to show that, for a force-free equilibrium, in flux coordinates the current density is proportional to the magnetic field through a coefficient which is a function of $r$ only, that is
\begin{equation}
\mu_0\mathbf{J}_0 = \sigma(r)\mathbf{B}_0.
\end{equation}

It is convenient to define, for a generic field $A$, its hatted version as $\hat{A}=\sqrt{g_w}A$, which hides the Jacobian contribution. The zeroth-order hatted magnetic field and current density components, function of $r$ only, are then:
\begin{equation}
\hat{B}_0^{\vartheta}=\Psi'_0 \qquad \hat{B}_0^{\phi}=F'_0 \qquad
\mu_0\hat{J}_0^{\vartheta}=\sigma\Psi'_0 \qquad \mu_0\hat{J}_0^{\phi}=\sigma F'_0.
\end{equation}

Given the $\sigma(r)$ profile, which is an input to the algorithm, the zeroth-order force balance yields the following equations:
\begin{eqnarray}
\frac{d}{dr}[K(r)\hat{B}_0^\phi] = -\sigma(r)\hat{B}_0^\vartheta \\
\frac{\partial}{\partial r}\left[\frac{g_{\vartheta\vartheta}^w}{\sqrt{g_w}}\hat{B}_0^\vartheta\right] -
\frac{\partial}{\partial\vartheta}\left(\frac{g_{r\vartheta}^w}{\sqrt{g_w}}\right)\hat{B}_0^\vartheta =
\sigma(r)\hat{B}_0^\phi.
\end{eqnarray}
The second equation contains metric coefficients which are function of $r$ and $\vartheta$. By using the expansion in harmonics described in Appendix B, it can be split into two equations, one for $\hat{B}_0^\vartheta$ and one for $\Delta$. Furthermore, because of the nonlinearity given by the fact that the metric coefficients depend on $\Delta$, it is convenient to introduce a perturbative expansion:
\begin{eqnarray}
\hat{B}_0^\vartheta = \hat{B}_1^\vartheta + \hat{B}_2^\vartheta + \ldots, \qquad
\hat{B}_2^\vartheta = o(\epsilon^2)\hat{B}_1^\vartheta \\
\hat{B}_0^\phi = \hat{B}_1^\phi + \hat{B}_2^\phi + \ldots, \qquad
\hat{B}_2^\phi = o(\epsilon^2)\hat{B}_1^\phi.
\end{eqnarray} 
The resulting equations for the lowest order contribution are 
\begin{eqnarray}
\frac{d}{dr}\left(\frac{R_0}{r}\hat{B}_1^\phi\right) = -\sigma(r)\hat{B}_1^\vartheta \\
\frac{d}{dr}\left(\frac{r}{R_0}\hat{B}_1^\vartheta\right) = \sigma(r)\hat{B}_1^\phi
\end{eqnarray}
which can be solved for $\hat{B}_1^\vartheta(r)$ and $\hat{B}_1^\phi(r)$. The solution starts from the magnetic axis, where regularity imposes $\hat{B}_1^\vartheta \sim \sigma_0 R_0 r/2$ and $\hat{B}_1^\phi \sim r$, and proceeds to the edge. The solutions are then rescaled so as to match a boundary condition, for example the edge poloidal field. 

These solutions are then be plugged into the equation for the shift
\begin{equation}
\Delta'' + \frac{\Delta'}{r}\left(1+2r\frac{d\hat{B}_1^\vartheta/dr}{\hat{B}_1^\vartheta}\right) 
+ \frac{1}{R_0} = 0.
\end{equation}
which is solved using the boundary conditions $\Delta'(0)=0$ and an assigned value for $\Delta(b)$, $b$ being the radius at which the magnetic measurements yielding the condition are located.

Having determined the shift $\Delta(r)$, the next order correction to the fields can be computed by
\begin{eqnarray}
\fl
\frac{d}{dr}\left(\frac{R_0}{r}\hat{B}^\phi_2\right) + 
\frac{d}{dr}\left[\frac{R_0}{r}\left(\frac{\Delta}{R_0}+\frac{r}{2R_0}\Delta'-\frac{r^2}{2R_0^2}\right)
\hat{B}_1^\phi\right] = -\sigma(r)\hat{B}^\vartheta_2 \\
\fl\frac{d}{dr}\left(\frac{r}{R_0}\hat{B}^\vartheta_2\right) + 
\frac{d}{dr}\left[\frac{r}{R_0}\left(\frac{r^2}{2R_0^2}+\frac{\Delta'^2}{2}+\frac{r}{2R_0}\Delta'
-\frac{\Delta}{R_0}\right)\hat{B}_1^\vartheta\right] = \sigma(r)\hat{B}^\phi_2.
\end{eqnarray}
Again, these equations are solved starting from the axis, where regularity requires $\hat{B}_2^\vartheta \sim 3r/(2\sigma_0 R_0)$ and $\hat{B}_2^\phi \sim [3/(\sigma_0^2R_0^2)-\Delta_0/R_0]r$, $\Delta_0$ being the shift of the magnetic axis. In practice this correction turns out to be very small.

It is worth noting that this method of computing the zeroth-order axisymmetric equilibrium, if compared to the standard Grad-Shafranov equation, has the advantage of requiring simply the solution of five ordinary differential equations. This is obtained at the price of being restricted to deal with circular flux surfaces, which is however reasonable for present day RFP devices.
While $\sigma(r)$ can in principle be any function, for the application described in the following the customary parametrization called $\alpha$-$\Theta_0$ model has been used, that is 
\begin{equation}
\sigma = \frac{2\Theta_0}{a}\left[1-\left(\frac{r}{a}\right)^\alpha\right].
\end{equation}
The two free parameters $\Theta_0$ and $\alpha$ are adjusted so as to obtain given values of the two dimensionless parameters $\Theta$ and $F$, which are the well known pinch and reversal parameters used to describe RFP plasmas.

\section{First-order contribution to equilibrium}
\label{first}

The next step is to add a non-axisymmetric perturbation to (\ref{B0}). In the gauge $A_r=0$, where $A_r$ is the covariant radial component of the vector potential $\mathbf{A}$, the total magnetic field can be written as
\begin{equation}
\mathbf{B} = \nabla F\times\nabla\vartheta - \nabla\Psi\times\nabla\phi
\label{B_perturb}
\end{equation}
where now $F$ and $\Psi$ depend on all three coordinates.
The $w^i = (r, \vartheta, \phi)$ coordinates 
are not any more flux coordinates for the perturbed magnetic field, which means that magnetic field lines of $\mathbf{B}$ are not straight in this coordinate system, and the potentials $F$ and $\Psi$ are not any more flux functions. These potentials, related to the vector potential covariant components, can be Fourier expanded as
\begin{equation}
\Psi(r,\vartheta,\phi) = -A_{\phi}(r,\vartheta,\phi) =  \Psi_0(r) +
\sum_{n\neq 0, m}\psi^{m,n}(r)e^{i(m\vartheta-n\phi)}
\end{equation}
\begin{equation}
F(r,\vartheta,\phi) = A_{\vartheta}(r,\vartheta,\phi) = F_0(r) +
\sum_{n\neq 0, m}f^{m,n}(r)e^{i(m\vartheta-n\phi)}
\end{equation}
The perturbed quantities contain $n \neq 0$ terms only, and the harmonics amplitudes are complex.

Given the representation (\ref{B_perturb}) of the magnetic field, the total hatted contravariant magnetic field components are
\begin{equation}
\hat{B}^\vartheta=\frac{\partial\Psi}{\partial r} \qquad
\hat{B}^\phi=\frac{\partial F}{\partial r} \qquad
\hat{b}^r=-\frac{\partial F}{\partial\phi}-\frac{\partial\Psi}{\partial\vartheta}.
\label{B_up_hatted}
\end{equation}
Computing the total current density components from Amp\'ere's law and plugging them into the first-order force balance equation
\begin{equation}
\mathbf{j}\times\mathbf{B}_0+\mathbf{J}_0\times\mathbf{b}=
\frac{1}{\sqrt{g_w}}\epsilon^{ijk}(\hat{j}^i\hat{B}_0^j+\hat{J}_0^i\hat{b}^j)\nabla w^k=0
\end{equation}
one obtains
the proportionality between perturbed radial current and perturbed radial magnetic field
\begin{equation}
\mu_0\hat{j}^r = \sigma(r)\hat{b}^r
\label{first1}
\end{equation}
and
\begin{equation}
\left(\frac{\partial}{\partial\vartheta}+q\frac{\partial}{\partial\phi}\right)
(\mu_0\hat{j}^\vartheta-\sigma\hat{b}^\vartheta)+\hat{b}^r\frac{d\sigma}{dr}=0
\label{first2}
\end{equation}
\begin{equation}
\left(\frac{\partial}{\partial\vartheta}+q\frac{\partial}{\partial\phi}\right)
(\mu_0\hat{j}^\phi-\sigma\hat{b}^\phi)+\hat{b}^rq\frac{d\sigma}{dr}=0.
\label{first3}
\end{equation}

By Fourier-transforming equations (\ref{first1}) and (\ref{first3}) (only two equations are needed, since for each mode there are two unknown functions $f^{m,n}$ and $\psi^{m,n}$) and using Amp\'ere's law the following equations are found:
\begin{eqnarray}
\fl
mK(r)\frac{df^{m,n}}{dr} +
n\left(\frac{g^w_{\vartheta\vartheta}}{\sqrt{g_w}}\right)^{0,0} \frac{d\psi^{m,n}}{dr} - 
\sigma(nf^{m,n}-m\psi^{m,n}) \nonumber \\
\fl
- in\left(\frac{g^w_{r\vartheta}}{\sqrt{g_w}}\right)^{1,0}
[nf^{m+1,n}-nf^{m-1,n}-(m+1)\psi^{m+1,n}+(m-1)\psi^{m-1,n}] \nonumber \\
\fl
+ n \left(\frac{g^w_{\vartheta\vartheta}}{\sqrt{g_w}}\right)^{1,0}
\left[\frac{d\psi^{m+1,n}}{dr}+\frac{d\psi^{m-1,n}}{dr}\right] = 0
\label{first4}
\end{eqnarray}

\begin{eqnarray}
\fl
\frac{d}{dr}\left(K(r)\frac{df^{m,n}}{dr}\right) +
\sigma\frac{d\psi^{m,n}}{dr} -
n\left(\frac{g^w_{rr}}{\sqrt{g_w}}\right)^{0,0}[nf^{m,n}-m\psi^{m,n}] -
\frac{nf^{m,n}-m\psi^{m,n}}{m-nq}\frac{d\sigma}{dr} \nonumber \\
\fl
- n\left(\frac{g^w_{rr}}{\sqrt{g_w}}\right)^{1,0}
[nf^{m+1,n}+nf^{m-1,n}-(m+1)\psi^{m+1,n}-(m-1)\psi^{m-1,n}] \nonumber \\
\fl
- in\left(\frac{g^w_{r\vartheta}}{\sqrt{g_w}}\right)^{1,0}
\left[\frac{d\psi^{m+1,n}}{dr}-\frac{d\psi^{m-1,n}}{dr}\right] = 0
\label{first5}
\end{eqnarray}

The method used for the solution of these equations is described in detail in ref. \cite{zt}. The solution requires the knowledge of the corresponding harmonic amplitude for the radial component of the magnetic field at some surface outside the plasma, which represents the boundary condition. Furthermore, if the mode has a resonant surface inside the plasma, a discontinuity in the eigenfunction derivative should be allowed. The magnitude of this discontinuity is obtained by imposing a further boundary condition, that is the amplitude of the toroidal magnetic field component at the same surface where the radial one is determined (that is the surface where the sensors are located).

\section{Helical flux surfaces}
\label{helical}

\begin{figure}[htb]
\centering
\includegraphics[width=0.8\columnwidth]{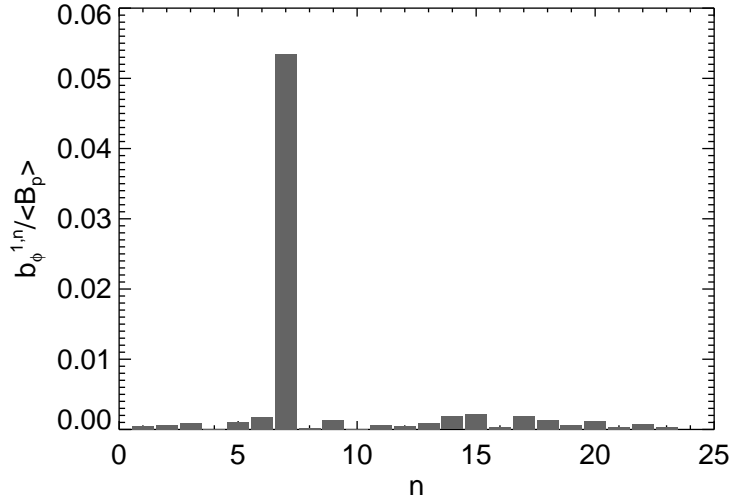}
\caption{$m=1$ mode spectrum in a 1.5 MA SHAx state in RFX-mod (shot 24598, t = 174 ms). The amplitudes of the Fourier modes on the toroidal field component, normalized to the average poloidal field, are shown.}
\label{fig_spectrum}
\end{figure}

As already mentioned in the introduction, the SHAx states will be modeled as pure Single Helicity states composed of the superposition of the zeroth-order axisymmetric equilibrium and of the dominant mode eigenfunction. The dominant mode, in the case of RFX-mod, is the m=1/n=7. 
An example of the $m=1$ mode spectrum for a 1.5 MA SHAx state obtained in RFX-mod is shown in Fig. \ref{fig_spectrum}. The spectrum gives the amplitude of the toroidal field component, normalized to the average poloidal field, measured outside the plasma. It can be seen how the $n=7$ mode clearly dominates over all the others, justifying the choice of taking it to be part of the equilibrium.

\begin{figure}[htb]
\centering
\includegraphics[width=0.8\columnwidth]{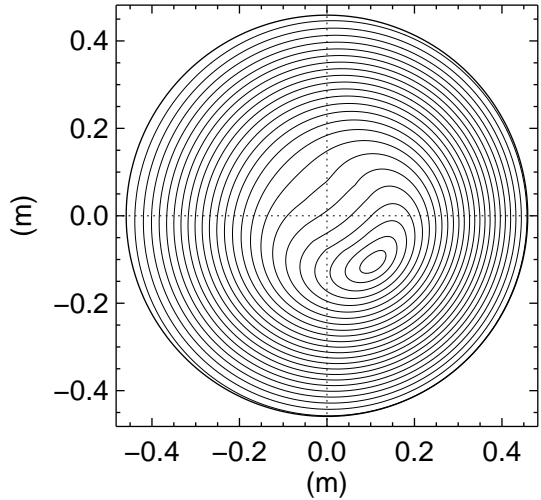}
\caption{Flux surfaces in a 1.5 MA SHAx state in RFX-mod (shot 24598, t = 174 ms), computed as contours of the helical flux $\chi$.}
\label{fig_surfaces}
\end{figure}

The poloidal and toroidal flux will be taken as:
\begin{equation}
\Psi(r,u) = \Psi_0(r) + \psi^{m,n}(r)e^{iu} + c.c.
\end{equation}
\begin{equation}
F(r,u) = F_0(r) + f^{m,n}(r)e^{iu} + c.c.
\end{equation}
where
\begin{equation}
u = m \vartheta - n \phi
\end{equation}
and $\Psi_0(r)$, $F_0(r)$, $\psi^{m,n}(r)$ and $f^{m,n}(r)$ are computed as described in the preceding sections.
The helical flux
\begin{eqnarray}
\chi(r,u) & = & m\Psi - nF = \nonumber \\
& = & m\Psi_0 - nF_0 + (m\psi^{m,n}-nf^{m,n}) \exp(iu) + c.c.
\label{chi}
\end{eqnarray}
is constant on the resulting flux surfaces. An example of such surfaces, obtained as contours of $\chi$, is shown in Fig. \ref{fig_surfaces}. It can be seen that only the inner surfaces are significantly distorted, and assume a bean-like shape. On the contrary, the outer ones retain a quasi-circular shape, with a shift induced by the presence of an $m=1$ component in the equilibrium.

SHAx states often display profiles of the kinetic plasma quantities which are non-symmetric with respect to the vacuum chamber axis. This is a direct consequence of the asymmetry of the helical flux surfaces displayed in Fig. \ref{fig_surfaces}. A direct test of the goodness of the equilibrium reconstruction is therefore a study of the profiles of such quantities when plotted as a function of a flux function. In the following we shall use as abscissa for such plots the effective radius defined as
\begin{equation}
\rho=\sqrt{\frac{\chi-\chi_0}{\chi_a-\chi_0}}
\end{equation}
where $\chi_0$ is the helical flux on the discharge helical axis and $\chi_a$ is the helical flux of the outermost surface. The $\rho$ variable ranges between 0 (on the helical axis) and 1.

\begin{figure}[htb]
\centering
\includegraphics[width=\columnwidth]{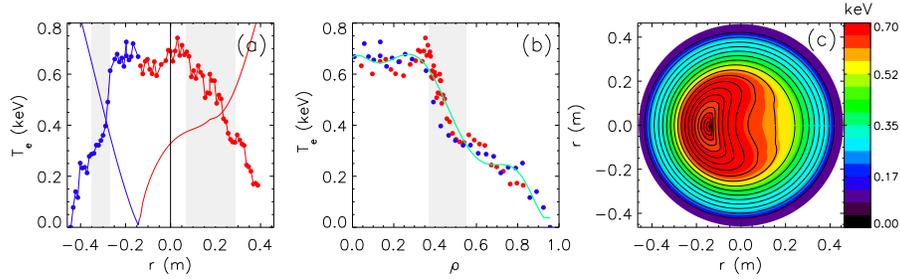}
\caption{(a) Electron temperature measured along a horizontal diameter of RFX-mod by the Thomson scattering technique. The two colours mark points that are on the two sides of the geometric chamber axis. (b) The same data of frame (a) plotted as a function of normalized the helical flux $\rho$. It can be seen how the two halves of the temperature profile collapse one onto the other. (c) Temperature map on the poloidal plane. The data refer to shot 24599 at $t=99$ ms.}
\label{fig_thomson}
\end{figure}

The first test of our equilibrium reconstruction is performed by considering an electron temperature ($T_e$) profile measured along a horizontal diameter of the chamber by an 84-point Thomson scattering system \cite{thomson}. An example of the measurements is shown in Fig. \ref{fig_thomson}a, where the previously mentioned asymmetry can be clearly seen. It is also observed, as already reported elsewhere \cite{carraro}, that the profile features strong electron transport barriers, revealed by the steep gradients in the shaded regions of the graph.  The same measurements, plotted as a function of $\rho$, are shown in Fig. \ref{fig_thomson}b. It is immediately apparent how the two half profiles, which in the previous plot were evidenced by the use of blue and red points, collapse one onto the other. This is a proof that the helical flux evaluated with our method is indeed a flux function, assuming that $T_e$ is (as reasonable given the very fast parallel thermal transport and the fact that the density profile is essentially flat). 
Fig. \ref{fig_thomson}c shows the resulting electron temperature map on the poloidal plane. The hot bean-shaped central region is clearly seen.

\begin{figure}[htb]
\centering
\includegraphics[width=\columnwidth]{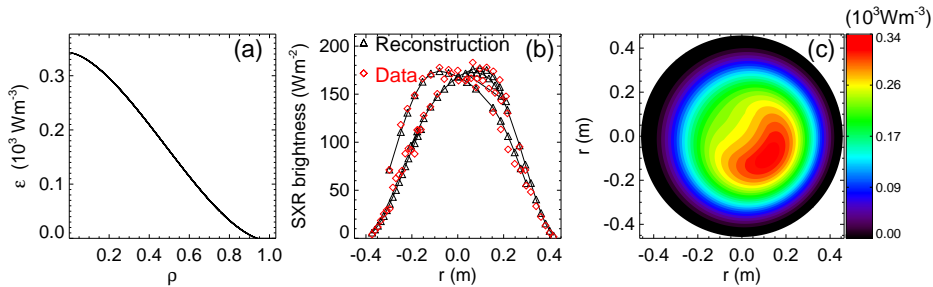}
\caption{(a) Soft X-ray emissivity profile given by the model of Equation (\ref{emissivity}), after parameter adjustment. (b)
Measured values of line-integrated soft X-ray emission (red diamonds) and values reconstructed according to the emissivity profile shown in panel a (black triangles). Emissivity pattern on the poloidal plane (c). The data refer to shot 26608 at $t=160$ ms.}
\label{fig_tomo}
\end{figure}

A second test has been performed using a 78-chord tomographic system, which measures the soft X-ray emission of the plasma \cite{tomo}. In this case, being the measurements integrated along lines of view, the emissivity $\epsilon$ has been assumed to be a function of $\rho$ only through a simple four-parameter model:
\begin{equation}
\epsilon(\rho)=\epsilon_0(1-\rho^\alpha)^\beta+\epsilon_1
\label{emissivity}
\end{equation}
The four parameters have been optimized so as to minimize the sum of the squared differences between the actual measurements and those reconstructed from the model. The final emissivity profile is shown in Fig. \ref{fig_tomo}a. The two classes of data, measured and reconstructed, are reproduced in Fig. \ref{fig_tomo}b. The figure displays two sets of points for each class, corresponding to the three vertical fans and to the single horizontal fan of measurement lines, respectively. It can be seen that the match is very good, despite the simplicity of the used model. The resulting emissivity pattern in the poloidal plane is shown in Fig. \ref{fig_tomo}c, showing once again the hot bean-shaped core region. 

\begin{figure}[htb]
\includegraphics[width=\columnwidth]{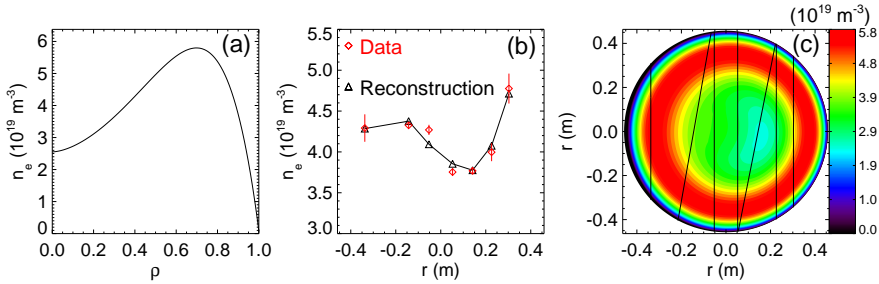}
\caption{(a) Plasma density profile given by the model of Equation (\ref{density}), after parameter adjustment.(b) Measured values of line-integrated density obtained through microwave reflectometry when a hydrogen pellet enters the bean-shaped plasma region (red diamonds) and values reconstructed according to the density profile shown in panel b (black triangles). Density pattern on the poloidal plans (c). The data refer to shot 24936 at $t=88$ ms.}
\label{fig_interf}
\end{figure}

Finally, line-averaged density measurements given by a 7-chord two-color interferometer \cite{interf} have been analyzed. Once again, a simple model, in this case with four parameters, has been assumed for the density dependence on $\rho$:
\begin{equation}
n(\rho) = n_0 - (n_0-n_1-n_{edge})\rho^\alpha - {n_1}\rho^\beta
\label{density}
\end{equation}
with $n_{edge}=10^{18}$ m$^{-3}$.
Usually, density profiles are flat, due to the fact that the discharge fueling is done by hydrogen released by the graphite first wall or puffed with valves. Thus, the density profile analysis has been performed on a case where a pellet had been injected into the plasma.
The density profile resulting from the optimization of the model parameters is shown in Fig. \ref{fig_interf}a, showing a hollow profile due to the effect of pellet ablation in the outer portion of the plasma. This is is different than the example shown in Ref.\cite{np}, where a peaked profile resulted from the pellet ablation taking place in the inner bean-shaped region. The measured values of line-integrated density are shown in \ref{fig_interf}b, together with the reconstructed values. A good agreement can be seen also in this case. Finally, the density pattern on the poloidal plane is reproduced in \ref{fig_interf}c. 

Summarizing, we have thus demonstrated, theoretically and experimentally, that the helical flux is a good flux function for SHAx states. Furthermore, the good mapping of the kinetic quantities on the computed flux surfaces shows that such surfaces are indeed, at least partially, found into the plasma. This is a proof of the emergence of some degree of order from the magnetic chaos present in MH condition.

\section{Flux surface averaging}
\label{averaging}
The computation of flux surface averages of different quantities is an essential step for building transport equations depending on one coordinate only (the flux surface label). The coordinate systems used up to this point are not appropriate for describing the helical flux surfaces, due to the fact that there are points which are not univocally identified by them. This can be understood by considering that for any inner flux surface, which does not contain the magnetic axis of the axisymmetric equilibrium, it is possible to identify two points which have the same value of the poloidal angles $\theta$ and $\vartheta$. We thus need to build a new coordinate system. This is done by adopting $\chi$ as radial coordinate, keeping $\phi$ as toroidal angle, and introducing a new poloidal angle, $\beta$, which rotates around the helical axis ($\nabla \chi =0$). The definition of $\beta$, with respect to the cylindrical coordinates, is
\begin{equation}
\beta=\tan^{-1}\frac{Z-Z_a(\phi)}{R-R_a(\phi)} ,
\label{defbeta}
\end{equation}
where $R_a(\phi)$ and $Z_a(\phi)$ represent the coordinates of the helical magnetic axis. We refer to Appendix C for the derivation of the metric tensor of the new $(\chi, \beta, \phi)$ coordinates.

The flux surface average of a generic quantity $A(\chi,\beta,\phi)$ in a toroidal system is given by \cite{dhaeseleer}
\begin{equation}
\langle A\rangle = \frac{\int\!\!\!\int d\beta d\phi \sqrt{g} A}
{\int\!\!\!\int d\beta d\phi \sqrt{g}} .
\end{equation}
The Jacobian $\sqrt{g}$ of the $(\chi, \beta, \phi)$ system can be computed relatively easily in terms of the Jacobian of the zeroth-order flux coordinates $w^i$ used above. One obtains
\begin{equation}
\frac{1}{\sqrt{g}} = \nabla\chi\cdot(\nabla\beta\times\nabla\phi) =
\left(
\frac{\partial\chi}{\partial r}\frac{\partial\beta}{\partial\vartheta}-
\frac{\partial\chi}{\partial\vartheta}\frac{\partial\beta}{\partial r}
\right) \frac{1}{\sqrt{g_w}}.
\label{jacob1}
\end{equation}
The derivatives appearing in expression (\ref{jacob1}) can be computed from the $\chi$ values given by Newcomb's equation and from (\ref{defbeta}), recalling (\ref{R}) and (\ref{Z}).

\begin{figure}[htb]
\begin{minipage} [htb] {0.5\columnwidth}
\centering
\includegraphics[width=\columnwidth]{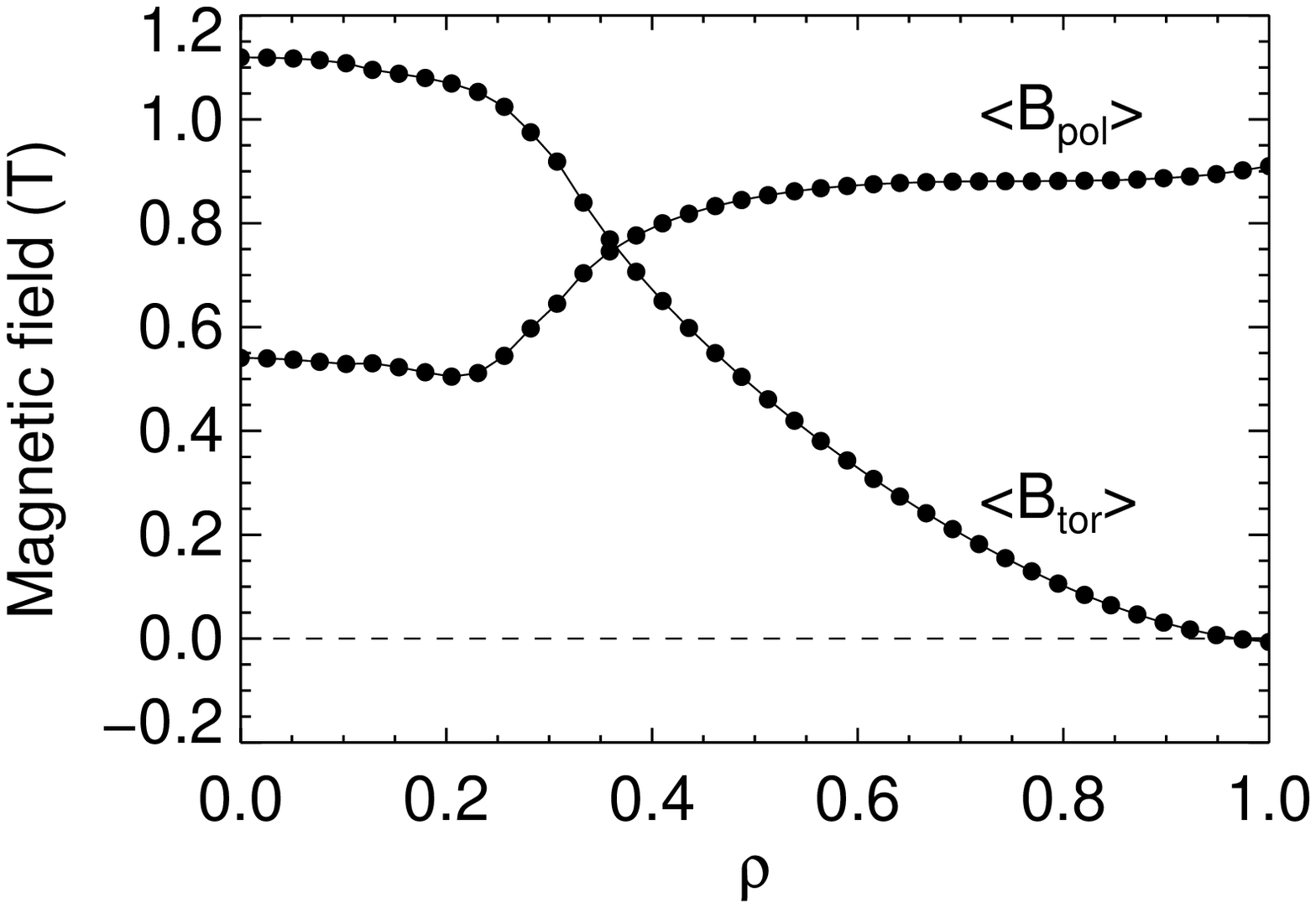}
\end{minipage}
\begin{minipage} [htb] {0.5\columnwidth}
\centering
\includegraphics[width=\columnwidth]{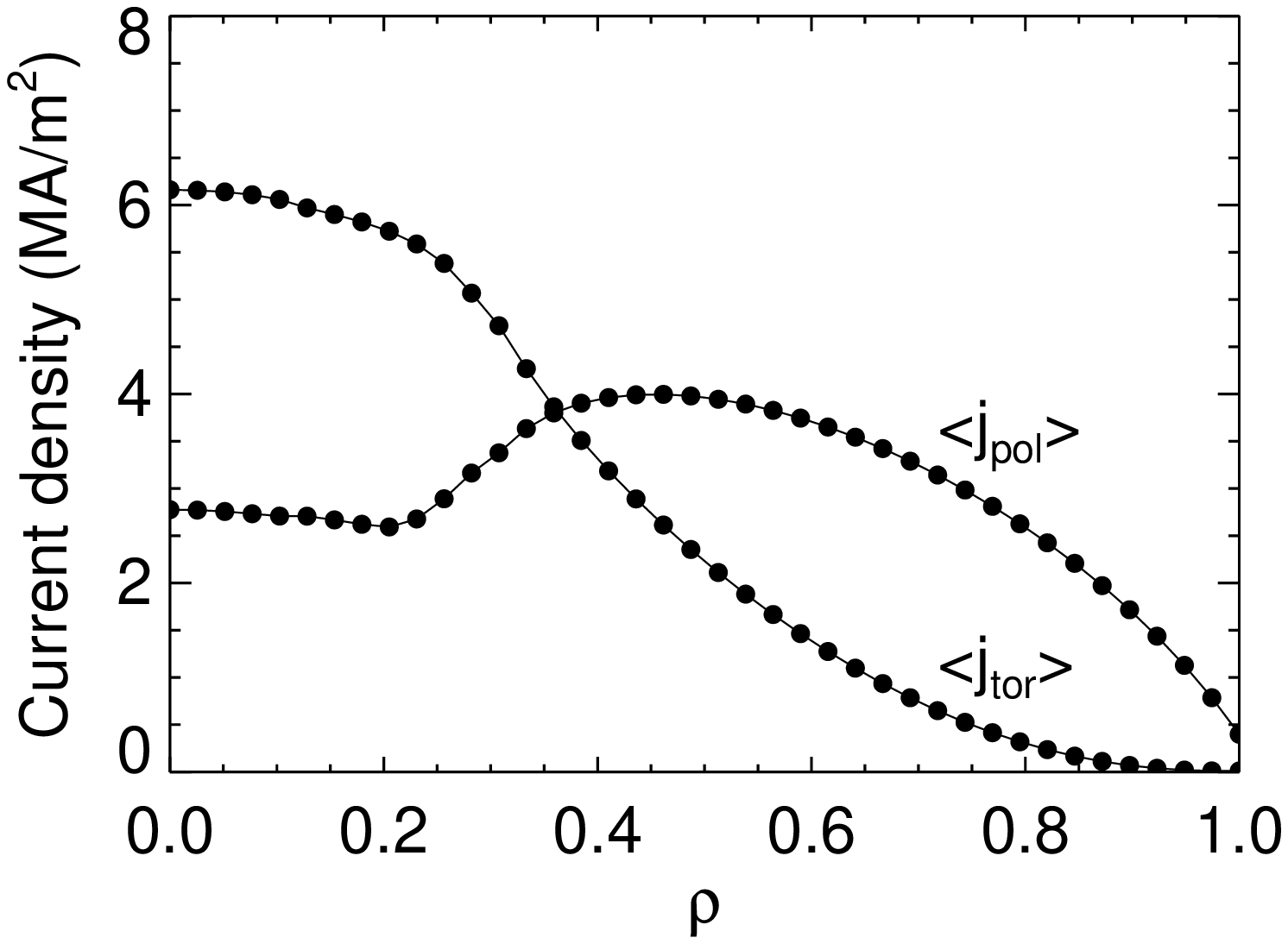}
\end{minipage}
\caption{Left: Flux surface averages of the toroidal and poloidal magnetic field components. Right: Flux surface averages of the toroidal and poloidal current density components. The data refer to shot 24598, at $t=174$ ms.}
\label{fig_averages}
\end{figure}

As an example of the application of this method, we show in Fig. \ref{fig_averages} the flux surface averages of the toroidal and poloidal covariant magnetic field and current density components. Using the relations that link the helical coordinates to the $w^i$ system of zeroth-order flux coordinates, the field components in the helical system can be obtained from those computed in the $w^i$ system. Equations (\ref{B_up_hatted}) return the hatted contravariant component $B^i$ of the magnetic field, so the metric contribution must be reintroduced: despite the assumption of single helicity in the fluxes, the contribution of the Jacobian harmonics (formula (\ref{g_expansion})) generates non-zero field components also for different mode numbers. This happens because of the toroidicity, which produces a toroidal coupling between modes with same toroidal mode number $n$ and different poloidal mode number $m$. The $B^i$ components result:
\begin{eqnarray}
B^r = i\left(nf^{1,n}(r)-m\psi^{1,n}(r) \right) G(r,\vartheta,\phi) + c.c. \\
\label{Br_up}
B^{\vartheta} = \left(\frac{1}{\sqrt{g_{w0}}}+\frac{2}{\sqrt{g_{w1}}}\cos\vartheta\right) \Psi'_0(r) + 
\psi'^{1,n}(r) G(r,\vartheta,\phi) + c.c.\\
\label{Bvartheta_up}
B^{\phi} = \left(\frac{1}{\sqrt{g_{w0}}}+\frac{2}{\sqrt{g_{w1}}}\cos\vartheta\right) F'_0(r) + 
f'^{1,n}(r) G(r,\vartheta,\phi) +c.c.
\label{Bphi_up}
\end{eqnarray}
with
\begin{equation}
\fl
G(r,\vartheta,\phi) = \frac{1}{\sqrt{g_{w0}}} e^{i(m\vartheta - n \phi)} +
\frac{1}{\sqrt{g_{w1}}} e^{i[(m-1)\vartheta-n\phi]} + 
\frac{1}{\sqrt{g_{w1}}} e^{i[(m+1)\vartheta-n\phi]}
\end{equation}
The definitions of $g_{w0}$ and $g_{w1}$ are given in Appendix B.

Writing
\begin{eqnarray}
\mathbf{B}= \underbrace{ B^r \, \mathbf{e_r} + B^\vartheta \, \mathbf{e_\vartheta} }_{\mathbf{B}_{pol}} +
\underbrace{ B^\phi \, \mathbf{e_\phi} }_{\mathbf{B}_{tor}} \; ,
\end{eqnarray}
the magnitudes of the poloidal and toroidal magnetic field components are:
\begin{equation}
B_{tor}=B_\phi=\sqrt{g_{\phi\phi}} B^\phi
\end{equation}
\begin{equation}
B_{pol}=\sqrt{g_{rr}(B^{r})^2+2g_{r\vartheta}B^rB^\vartheta+g_{\vartheta\vartheta}(B^{\vartheta})^2}.
\end{equation}
$B_{tor}$ can also be identified with the measured covariant components $B_\phi$ on the normalized basis vectors $\mathbf{e_\phi}$. 

The graphs of Fig. \ref{fig_averages} show that the flux surface average of the toroidal field component is monotonically decreasing, as for the standard cylindrical models of the RFP fields, with a maximum of 1.1 T which is now located on the helical axis, a slow decrease in the central part of the plasma, a knee around $\rho=0.25$, and a reversal in the outer part of the plasma. The poloidal component is also rather flat on the inner surfaces, where it has an almost uniform value around 0.5 T. The same features are displayed by the current density components, not surprisingly since this is a force-free equilibrium. The maximum toroidal current density, on the helical axis, is around 6 MA/m$^2$, while in the inner part of the plasma the poloidal component takes values a little larger than 2.5 MA/m$^2$. It can be also remarked that for $\rho>0.8$ the toroidal current density is negligible.

\begin{figure}[htb]
\begin{minipage} [htb] {0.5\columnwidth}
\centering
\includegraphics[width=\columnwidth]{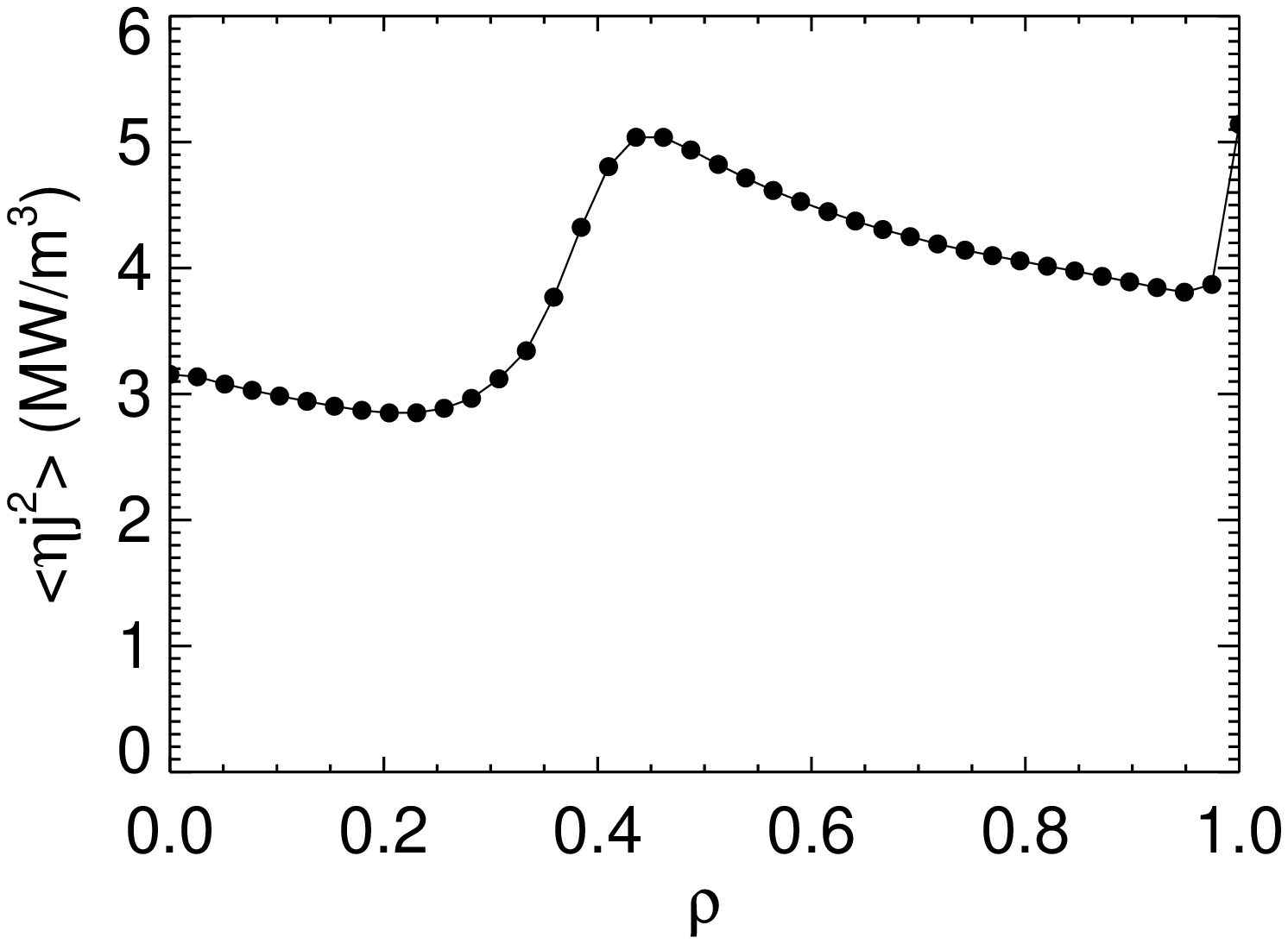}
\end{minipage}
\begin{minipage} [htb] {0.5\columnwidth}
\centering
\includegraphics[width=\columnwidth]{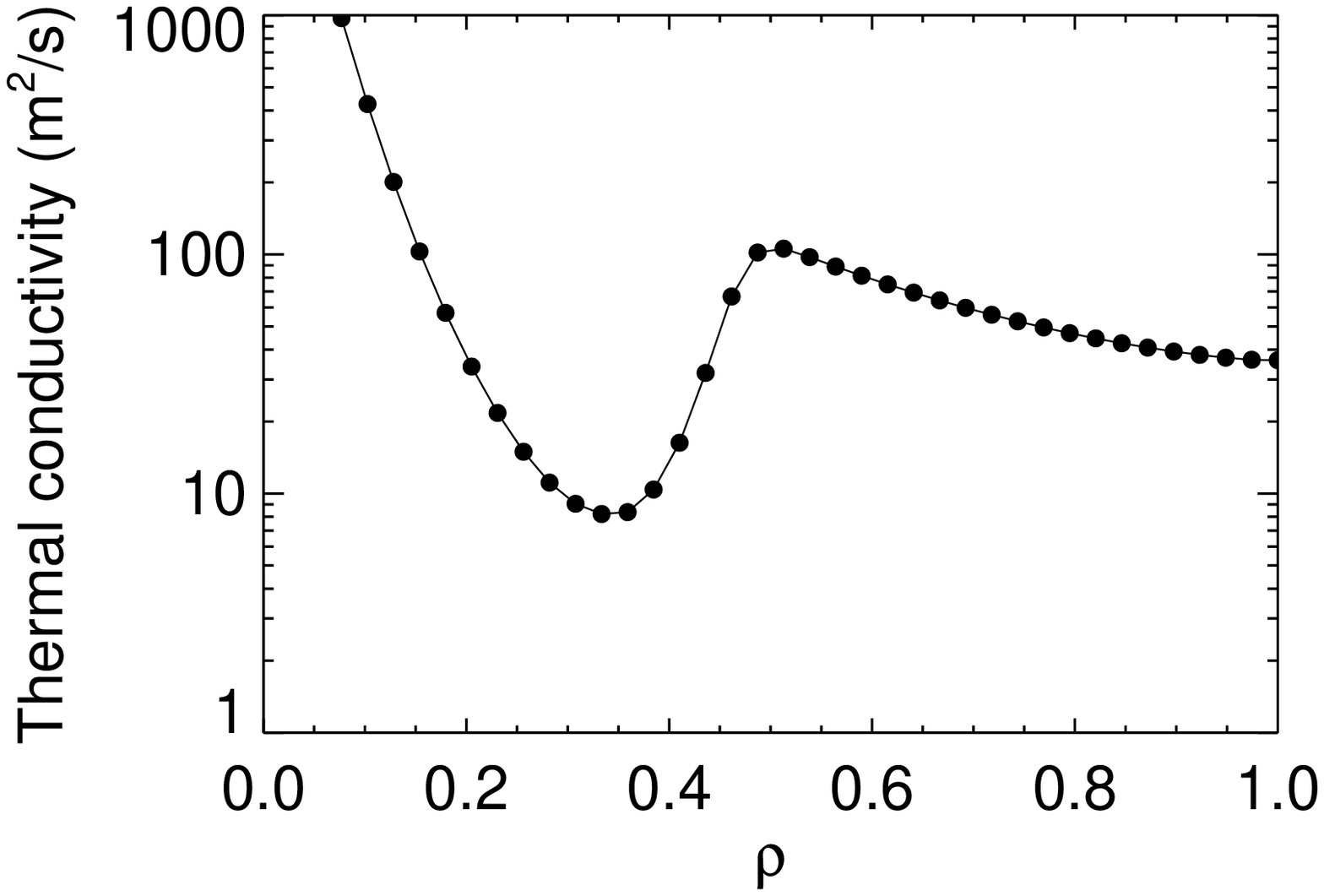}
\end{minipage}
\caption{Left: Flux surface average of dissipated ohmic power. Right: Thermal conductivity profile computed from the surface-averaged power balance. The data refer to shot 22182, at $t=49$ ms.}
\label{fig_power}
\end{figure}

Using the same approach to flux surface averaging, one can compute the average ohmic power, $\langle\eta J^2\rangle$. In doing this the Spitzer-H\"arm resistivity formula has been used, with the electron temperature profile measured by the Thomson scattering system. Furthermore, we have assumed a uniform effective charge profile, and the correction factor with respect to the $Z=1$ resistivity has been computed comparing the volume integral of the dissipated power to the actual input power, $P=VI$, assuming stationary conditions. The result is shown in the left panel of Fig. \ref{fig_power}. For this case, the correction factor turned out to be equal to 1.75, which appears a rather reasonable value. It is interesting to notice how the dissipated power displays a peak outside of the bean-shaped inner region. It should however be emphasized that this feature crucially depends on the profile chosen for $\sigma(r)$.

The averaged dissipated power can be used in a simplified power balance equation,
\begin{equation}
\frac{1}{V'}\frac{\partial}{\partial\chi}\left(V' \langle g^{11}\rangle \kappa n\frac{dT}{d\chi}\right)=\langle\eta J^2\rangle
\end{equation}
to compute the thermal conductivity $\kappa$. Here $g^{11}=\nabla\chi\cdot\nabla\chi$ is the first metric tensor element, which has been computed explicitly in Appendix C, while $V'=dV/d\chi$ is the specific volume. Using the temperature profile given in Fig. \ref{fig_thomson} and the averaged power of Fig. \ref{fig_power} (left), the thermal conductivity profile shown in the right panel of Fig. \ref{fig_power} is obtained. It can be seen that the thermal conductivity displays a minimum, at a value around 8 m$^2$/s, corresponding to the strong gradient in the temperature profile. This is an order of magnitude lower than values obtained in MH conditions \cite{bartiromo}.

\section{Ohmic constraint}
\label{ohmic}

As in the RFP there is a large current flowing in the plasma, a natural question which arises when an ideal equilibrium has been computed is whether such current is consistent with Ohm's law. In MH states this cannot be easily assessed, since Ohm's law includes a ``dynamo electric field'' which results from the effect of perturbations in the quadratic $\mathbf{v}\times\mathbf{B}$ term. In SH states, on the other hand, one would require Ohm's law to be valid for the helical equilibrium. 

The parallel component of Ohm's law gives, in stationary conditions,
\begin{equation}
-\nabla\cdot\left(\Phi\mathbf{B}\right) + \frac{V_t}{2\pi}\mathbf{B}\cdot\nabla\phi =
\eta\mathbf{j}\cdot\mathbf{B}
\end{equation}
where $\Phi$ is the electrostatic potential and $V_t$ is the toroidal loop voltage. 
Multiplying by the Jacobian of the coordinate system in use, and integrating over the two angular variables, the term involving the electrostatic potential cancels, and one finds
\begin{equation}
\frac{V_t}{2\pi} \langle B^\phi\rangle = \eta\langle\mathbf{j}\cdot\mathbf{B}\rangle.
\label{oh_constraint}
\end{equation}
This relationship defines the Ohmic constraint that any stationary equilibrium should satisfy. 

\begin{figure}[htb]
\centering
\includegraphics[width=0.6\columnwidth]{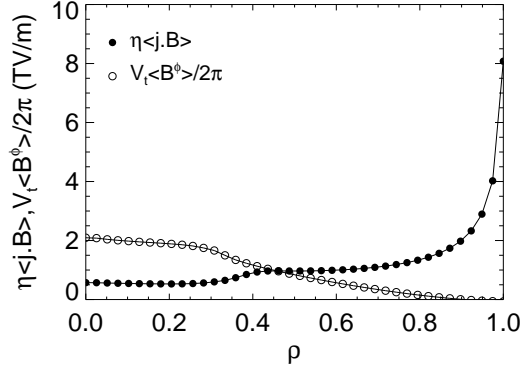}
\caption{Left hand side (open circles) and right hand side (solid circles) of equation (\ref{oh_constraint}), plotted as a function of the effective radius $\rho$. The data refer to shot 22182, at $t=49$ ms.}
\label{fig_ohmic}
\end{figure}

The two sides of the Ohmic constraint are plotted in Fig. \ref{fig_ohmic} for a typical 1.5 MA SHAx state. As for the power balance described in the preceding section, a flat $Z_{eff}$ profile has been assumed. A remarkable discrepancy can be seen, with the first term being larger than the second one in the inner portion of the plasma, and smaller in the outer one. Such discrepancy could be partially resolved assuming a profile of effective charge with a peak in the center of the plasma, instead of the flat profile which has been assumed here. Even though a screening of impurity influx has been observed in Laser Blow-Off experiments \cite{LBO}, a trapping in the central part of the plasma of impurities existing prior to the SHAx state onset might occur. Even if this is the case, the $\alpha-\Theta_0$ model assumed for the zeroth-order parallel current density appears anyway to be not appropriate, since in the outer part of the plasma the profile of $\langle B^\phi \rangle$ changes sign, whereas the other one does not. This requires either to assume a profile of $\sigma$ changing sign in this region, or a residual dynamo contribution of the secondary modes. These considerations point to the need of performing equilibrium calculations which take into account the Ohmic constraint.

\section{Conclusions}
\label{conclusions}
In this paper we have described a perturbative method for the computation of helical equilibria for the SHAx states of RFP plasmas in force-free approximation. By using the limitation of assuming a zeroth-order axisymmetric equilibrium with circular flux surfaces, it has been possible to reduce its calculation, and the subsequent calculation of the dominant mode eigenfunction, to the solution of ordinary differential equations. The helical flux given by the superposition of the zeroth-order equilibrium and of the eigenfunction has been demonstrated to be a good flux function, by comparison with spatially resolved measurements of different plasma quantities. Both the force free approximation and the perturbative approach do not appear to give errors which go beyond the statistical uncertainty of the measurements used for the validation, at least for the conditions which have been explored up to now in the RFX-mod device (plasma currents up to 1.8 MA). In particular, being the $\beta$ value of these plasmas of a few percent, the approximation of neglecting the pressure gradient term in the equilibrium equation appears to be valid, despite the presence of a region of strong gradient. Large increases in performance are allowed before this approximation breaks down.

As already discussed in ref. \cite{np}, the possibility of mapping the kinetic measurements on the helical flux surfaces resulting from the equilibrium reconstruction constitutes a proof of the existence of at least some reminiscence of such surfaces in the plasma. While there are no proofs that actual KAM surfaces are formed, it has been shown that cantori, that is broken KAM surfaces, are sufficient for supporting temperature gradients \cite{hudson}. These ordered structures have been named ``ghost surfaces'' \cite{gole}. The possibility that such ghost surfaces appear in the otherwise chaotic RFX plasma when the SHAx state is attained is the most likely explanation for the experimental observations.

The availability of an equilibrium reconstruction for the high performance SHAx states, as given by the SHEq code which implements the method that we have described, opens the path for several lines of research. Among the work in progress are the use of the ASTRA code \cite{astra} for performing transport calculations taking into account the helical geometry and the evaluation of crucial quantities such as the safety factor profile. The output of our equilibrium calculation could also serve as the basis for stability calculations or for the evaluation of the neoclassical transport coefficients in SHAx plasmas.

\ack
This work was supported by the European Communities under the contract of Association between EURATOM/ENEA. The authors would like to acknowledge useful discussions with A.~H.~Boozer, D.~F.~Escande and L.~Marrelli.

\appendix

\section*{Appendix A}
\setcounter{section}{1}
The covariant metric tensor for a curvilinear coordinate system $u^i=(u^1, u^2, u^3)$ is defined by
\begin{equation}
g_{ij}^u=\mathbf{e}_i \cdot \mathbf{e}_j \, ,
\label{g_down}
\end{equation}
where
\begin{eqnarray}
\mathbf{e}_i = \frac{\partial \mathbf{x}}{\partial u^i} \, ,
\label{base_down}
\end{eqnarray}
while the contravariant one is defined by
\begin{equation}
g_u^{ij}=\nabla u^i \cdot \nabla u^j
\label{g_up}
\end{equation}
where
\begin{equation}
\nabla u^i = \frac{\partial u^i}{\partial \mathbf{x}} \, .
\label{base_up}
\end{equation}

The two tensors are related by $g_u^{ij}\cdot g_{jk}^u=\delta_k^i$, so $g_u^{ij}$ is the inverse matrix of $g_{jk}^u$.
The Jacobian of the coordinate system is
\begin{equation}
\sqrt{g_u} = \sqrt{\det[g_{ij}^u]} = (\nabla u^1 \cdot \nabla u^2 \times \nabla u^3)^{-1}
\end{equation}
Given a vector $\mathbf{A}$, its contravariant components are defined as
\begin{equation}
A^i = \mathbf{A}\cdot\nabla u^i \qquad 	\mathrm{or} \qquad  A^i=g_u^{ij} A_j
\end{equation}
the second one expresses the contravariant component in terms of the covariant ones. In the same way,
\begin{equation}
A_i = \mathbf{A}\cdot \mathbf{e}_i \qquad \mathrm{or} \qquad A_i=g^u_{ij} A^j
\end{equation}

\section*{Appendix B}
\setcounter{section}{2}

The geometric coordinates $u^i=(r,\theta,\phi)$ are defined by relationships (\ref{geo1}) and (\ref{geo2}).
Using (\ref{g_down}) and (\ref{base_down}) the $g_{ij}^u$ elements, that link them to the cartesian coordinates, can be explicitly calculated:
\begin{eqnarray}
g_{ij}^{u} =
\left( \begin{array}{ccc}
1-2 \Delta' \cos \theta + \Delta'^2   &   r \Delta' \sin \theta   &    0 \\
r \Delta' \sin \theta                 &   r^2                     &    0 \\
0                                     &  0                        & R^2 \\
\end{array} \right)
\end{eqnarray}
The $g^{ij}_u$ elements of the inverse matrix are usually calculated from the formulas for the inversion of a block diagonal matrix ($g_{r\phi}^u=g_{\phi r}^u=0$ and $g_{\vartheta \phi}^u=g_{\phi \vartheta}^u=0$). The Jacobian $\sqrt{g_u} = \sqrt{\det[g_{ij}^u]} = (\nabla r \cdot \nabla \theta \times \nabla \phi)^{-1}$ of the geometric coordinates is given by
\begin{equation}
\sqrt{g_u} = r R (1-\Delta' \cos \theta)
\end{equation}

In the text we have defined the coordinate system $w^i=(r, \vartheta, \phi)$, which is the flux coordinate system of the zeroth-order axisymmetric equilibrium $\mathbf{B}_0$, as the coordinate system that one obtains deforming the poloidal angle $\theta$ of the geometrical coordinates in order to achieve straight magnetic field lines:
\begin{equation}
\vartheta= \theta + \lambda(r, \theta)
\end{equation}
From Amp\`ere's law we have also obtained an explicit expression for $\lambda(r, \theta)$, (\ref{lambda}).
The contravariant metric tensor elements can be computed using the relations between the gradients of the two coordinate systems:
\begin{eqnarray}
\nabla r = \nabla r \\
\label{nablavartheta}
\nabla \vartheta = \left(1+ \frac{\partial \lambda}{\partial \theta} \right) \nabla \theta 
                     + \frac{\partial \lambda}{\partial r} \nabla r \\
\nabla \phi = \nabla \phi
\end{eqnarray}
It is therefore found
\begin{eqnarray}
g_w^{r\vartheta} = g_w^{\vartheta r} = \frac{\partial\lambda}{\partial r}g^{rr}_u + \left(\frac{\partial\lambda}{\partial\theta}+1\right)g^{r\theta}_u \\
g_w^{\vartheta\vartheta} = \left(\frac{\partial\lambda}{\partial r}\right)^2 g^{rr}_u + 2\left(\frac{\partial\lambda}{\partial\theta}+1\right)\frac{\partial\lambda}{\partial r}g^{r\theta}_u +
\left(\frac{\partial\lambda}{\partial\theta}+1\right)^2 g^{\theta\theta}_u
\end{eqnarray}
while all the other elements are equal to those of the geometric coordinates.
The $g_{ij}^w$ elements of the inverse matrix can again be calculated from the formulas for the inversion of a block diagonal matrix ($g^{r\phi}_w=g^{\phi r}_w=0$ and $g^{\vartheta \phi}_w=g^{\phi \vartheta}_w=0$), or can be computed by writing the relation between the $w^i$ and the $(R,\phi,Z)$ system. In this second case, using equations (\ref{R}) and (\ref{Z}) the covariant metric tensor elements are found to be
\begin{eqnarray}
\fl
g_{rr}^w = 1 + \left(2\Delta'^2+\frac{r^2}{2R_0^2}+\frac{r^2}{2}\Delta''^2-\frac{r^2}{R_0}\Delta''-
\frac{2r}{R_0}\Delta'+r\Delta'\Delta''\right) - 2\Delta'\cos\vartheta + o(\epsilon^2) \\
\fl 
g_{\vartheta\vartheta}^w = r^2 \left(1+\frac{r^2}{2R_0^2}+\frac{1}{2}\Delta'^2-
\frac{r}{R_0}\Delta'\right) - 2r^2\left(\frac{r}{R_0}-\Delta'\right)\cos\vartheta + o(\epsilon^2b^2) \\
\fl
g_{r\vartheta}^w = r \left(r\Delta''+\Delta'-\frac{r}{R_0}\right)\sin\vartheta + o(\epsilon^2b) \\
\fl
g_{\phi\phi}^w = R^2.
\end{eqnarray}
with $g_{r\phi}^w=g_{\phi r}^w=g_{\vartheta\phi}^w=g_{\phi\vartheta}^w=0$.
In computing these elements the approximation used in ref.\cite{fitz} has been adopted of retaining the secular terms (i.e. those not dependent on $\vartheta$) up to $o(\epsilon^2)$ and the harmonics up to $o(\epsilon)$.

The Jacobian is
\begin{equation}
\frac{1}{\sqrt{g_w}} = \frac{K(r)}{R^2}
\end{equation}
with $K(r)$ already written in (\ref{K_r}). In order to compute the harmonics of the first order perturbation, one wishes to expand also the metric tensor elements:
\begin{equation}
\frac{1}{\sqrt{g_w}} = \frac{1}{\sqrt{g_{w0}}} + \frac{1}{\sqrt{g_{w1}}}e^{i \vartheta} + c.c.
\label{g_expansion}
\end{equation}
with
\begin{eqnarray}
\frac{1}{\sqrt{g_{w0}}} = \frac{1}{r R_0} \left( 1+ \frac{2 r^2}{R_0^2} - \frac{\Delta}{R_0} - \frac{r}{2 R_0} \Delta' + o(\epsilon^3) \right) \\
\frac{1}{\sqrt{g_{w1}}} = \frac{1}{\sqrt{g_{w0}}} \left( \frac{r}{R_0} + o (\epsilon^3) \right) ,
\end{eqnarray}
being the $m=0/n=0$ and $m=\pm 1/n=0$ harmonics of the Jacobian respectively.

We can also relate the Jacobian $\sqrt{g_w}$ to $\sqrt{g_u}$. Using (\ref{nablavartheta}) one obtains
\begin{eqnarray}
\frac{1}{\sqrt{g_w}}=\frac{1}{\sqrt{g_u}}\left(1+\frac{\partial\lambda}{\partial\theta}\right).
\end{eqnarray}

Finally, the metric tensor element combinations appearing in eqs. (\ref{first4}) and (\ref{first5}) are
\begin{eqnarray}
\fl
\left(\frac{g_{rr}^w}{\sqrt{g_w}}\right)^{0,0} =
\frac{1}{rR_0}\left[1+\Delta'^2\left(\frac{3}{2}+2r^2\left(\frac{\Psi''_0}{\Psi'_0}\right)^2\right)+
\frac{r}{R_0}\Delta'\left(4r\frac{\Psi''_0}{\Psi'_0}-\frac{7}{2}\right) +4\frac{r^2}{R_0^2}-\frac{\Delta}{R_0}+o(\epsilon^3)\right] \\
\fl
\left(\frac{g_{\vartheta\vartheta}^w}{\sqrt{g_w}}\right)^{0,0} =
\frac{r}{R_0}\left(1+\frac{r^2}{2R_0^2}+\frac{\Delta'^2}{2}+\frac{r}{2R_0}\Delta'-\frac{\Delta}{R_0}+
o(\epsilon^3)\right) \\
\fl
\left(\frac{g_{r\vartheta}^w}{\sqrt{g_w}}\right)^{0,0} = 
\frac{o(\epsilon^4b)}{\sqrt{g_{w0}}} \\
\fl
\left(\frac{g_{rr}^w}{\sqrt{g_w}}\right)^{\pm 1,0} = 
\frac{1}{\sqrt{g_{w0}}}\left(\frac{r}{R_0}-\Delta'+o(\epsilon^3)\right) \\
\fl
\left(\frac{g_{\vartheta\vartheta}^w}{\sqrt{g_w}}\right)^{\pm 1,0} = 
\frac{1}{\sqrt{g_{w0}}}\left(r^2\Delta'+o(\epsilon^3b^2)\right) \\
\fl
\left(\frac{g_{r\vartheta}^w}{\sqrt{g_w}}\right)^{\pm 1,0} = 
\pm \frac{r}{2i\sqrt{g_{w0}}}\left(r\Delta''+\Delta'-\frac{r}{R_0}+o(\epsilon^3)\right).
\end{eqnarray}

\appendix
\section*{Appendix C}
\setcounter{section}{3}

The  definition of the helical coordinates $(\chi, \beta, \phi)$ used in this paper to describe SHAx states is given in the text: the radial coordinate is the helical flux $\chi$ (\ref{chi}), the poloidal-like angle $\beta$ is given in (\ref{defbeta}), and the toroidal angle $\phi$ is the usual one.

The contravariant metric tensor elements are derived from the relation between the gradients in this coordinate system and in the flux coordinate system of the zeroth-order axisymmetric magnetic field, $w^i=(r, \vartheta, \phi)$.
By definition the $\beta$ angle is a function of the cylindrical $(R, \phi, Z)$ coordinates, where the toroidal dependence is due to the dependence on the toroidal angle of the magnetic axis position, $R_a(\phi)$ and $Z_a(\phi)$, so that
\begin{eqnarray}
\nabla \chi = \frac{\partial \chi}{\partial r} \nabla r +
              \frac{\partial \chi}{\partial \vartheta} \nabla \vartheta +
              \frac{\partial \chi}{\partial \phi} \nabla \phi \\
\nabla \beta = \frac{\partial \beta}{\partial r} \nabla r +
               \frac{\partial \beta}{\partial \vartheta} \nabla \vartheta +
               \frac{\partial \beta}{\partial \phi} \nabla \phi \\
\nabla \phi = \nabla \phi 
\label{gbeta_up}
\end{eqnarray}
These relationships allow to compute all the components of the contravariant metric tensor. Here we write explicitly just the $g^{\chi \chi}$ element (used in the text for the calculation of the thermal conductivity):
\begin{eqnarray}
\fl
g^{\chi \chi}=\nabla \chi \cdot \nabla \chi
             = \left( \frac{\partial \chi}{\partial r} \nabla r +
               \frac{\partial \chi}{\partial \vartheta} \nabla \vartheta +
               \frac{\partial \chi}{\partial \phi} \nabla \phi \right) \cdot
	             \left( \frac{\partial \chi}{\partial r} \nabla r +
               \frac{\partial \chi}{\partial \vartheta} \nabla \vartheta +
               \frac{\partial \chi}{\partial \phi} \nabla \phi \right) =  \nonumber\\
\nonumber\\
  = \left( \frac{\partial \chi}{\partial r} \right)^2 g_w^{rr}
  + \left( \frac{\partial \chi}{\partial \vartheta} \right)^2 g_w^{\vartheta\vartheta}
  + \left( \frac{\partial \chi}{\partial \phi} \right)^2 g_w^{\phi \phi}
  + \left( \frac{\partial \chi}{\partial r} \right) \left( \frac{\partial \chi}{\partial \vartheta} \right)  g_w^{r \vartheta} \nonumber
\end{eqnarray}
The partial derivatives of $\chi$ with respect to $\vartheta$ and $\phi$ are straightforward to compute, whereas the one with respect to $r$ is evaluated through expressions (\ref{first4}) and (\ref{first5}).

The Jacobian is also derived from (\ref{gbeta_up}), following the definition
\begin{eqnarray}
\sqrt{g}=(\nabla \chi \cdot \nabla \beta \times \nabla \phi )^{-1},
\end{eqnarray}
so that
\begin{equation}
\frac{1}{\sqrt{g}} = \nabla\chi\cdot(\nabla\beta\times\nabla\phi) =
\left(
\frac{\partial\chi}{\partial r}\frac{\partial\beta}{\partial\vartheta}-
\frac{\partial\chi}{\partial\vartheta}\frac{\partial\beta}{\partial r}
\right) \frac{1}{\sqrt{g_w}}.
\label{jacob2}
\end{equation}
The proof of this simple relation goes as follows:
\begin{eqnarray}
\nabla \beta \times \nabla \phi  =  
\left(\frac{\partial \beta}{\partial r} \nabla r +
      \frac{\partial \beta}{\partial \vartheta} \nabla \vartheta +
      \frac{\partial \beta}{\partial \phi} \nabla \phi
		  \right) \times \nabla \phi = \nonumber \\
 = \frac{\partial \beta}{\partial r} \nabla r \times \nabla \phi +
     \frac{\partial \beta}{\partial \vartheta} \nabla \vartheta \times \nabla \phi \nonumber
\end{eqnarray}
so
\begin{eqnarray}
\nabla \chi \cdot \nabla \beta \times \nabla \phi  = \nonumber \\
= \left( \frac{\partial \chi}{\partial r} \nabla r +
              \frac{\partial \chi}{\partial \vartheta} \nabla \vartheta +
               \frac{\partial \chi}{\partial \phi} \nabla phi \right) \cdot
     \left(\frac{\partial \beta}{\partial r} \nabla r \times \nabla \phi +
           \frac{\partial \beta}{\partial \vartheta} \nabla \vartheta \times \nabla \phi \right) = \nonumber \\
= \frac{\partial \chi}{\partial r} \frac{\partial \beta}{\partial \vartheta}  \nabla r \cdot \nabla \vartheta \times \nabla \phi + \frac{\partial \chi}{\partial \vartheta} \frac{\partial \beta}{\partial r} \nabla \vartheta \cdot \nabla r \times \nabla \phi = \nonumber\\
=\left(\frac{\partial \chi}{\partial r} \frac{\partial \beta}{\partial \vartheta} -
		     \frac{\partial \chi}{\partial \vartheta} \frac{\partial \beta}{\partial r}
		     \right) \nabla r \cdot \nabla \vartheta \times \nabla \phi \nonumber
\end{eqnarray}
that is exactly relation (\ref{jacob2}). 
The partial derivatives of $\beta$ are computed using the definition of $\beta$ in terms of $R$ and $Z$ and relations (\ref{R}) and (\ref{Z}).

\section*{References}

\end{document}